\documentstyle[eqsecnum,aps]{revtex}
\begin{document}
\draft
\title{On the theory of the relativistic cross sections for stimulated
bremsstrahlung on an arbitrary electrostatic potential in the strong
electromagnetic field.}
\author{T. R. Hovhannisyan, A.G. Markossian, and G.F. Mkrtchian}
\address{Plasma Physics Laboratory, Yerevan State University, 1 A. Manukian, 375049\\
Yerevan, Armenia\\
E-mail: tesakfiz@sun.ysu.am}
\maketitle

\begin{abstract}
On the base of relativistic generalized eikonal approximation wave function
the multiphoton cross sections of a Dirac particle bremsstrahlung on an
arbitrary electrostatic potential and strong laser radiation field are
presented. In the limit of the Born approximation the ultimate analytical
formulas for arbitrary polarization of electromagnetic wave have been
obtained.
\end{abstract}


\section{Introduction}

In the previous papers \cite{1},\cite{2} so called generalized eikonal
approximation (GEA) has been developed in the relativistic quantum theory of
elastic scattering of Dirac particle on an arbitrary electrostatic field (%
\cite{1}) and for the stimulated bremsstrahlung (SB) in the presence of an
external strong electromagnetic (EM) radiation field (\cite{2}). These wave
functions enable to leave the framework of ordinary eikonal approximation as
for elastic as well as for inelastic scattering (\cite{3} and \cite{4}),
that is not applicable beyond the interaction region ( $z\ll pa^2/\hbar $,
where $z$ is the coordinate along the direction of initial momentum $%
\overrightarrow{p}$ of the particle and $a$ is the characteristic size of
the interaction region). It means, that such a wave function is applicable
in both quantum and quasiclassical limits, i.e., connects the particle wave
functions of the Born and ordinary eikonal approximations.

The first nonrelativistic treatment of SB in the Born approximation has been
carried analytically by Bunkin and Fedorov (\cite{3}) and then that approach
has been extended to the relativistic domain by Denisov and Fedorov (\cite{4}%
). Further the multiphoton cross sections of SB process have been obtained
in the low-frequency \cite{5} and eikonal \cite{6} approximations. In the
present paper the relativistic cross sections of SB in the scope of the
above-mentioned GEA approximation are obtained and comparably simple
formulas for the transition amplitudes and cross sections for the
Dirac-particle scattering in the presence of an arbitrary polarized plane
electromagnetic wave in the limit of the Born approximation are obtained.
Note, that recently (\cite{7}) the complicate formulas for the relativistic
cross sections of SB in the case of circularly polarized monochromatic wave
corresponding to the relativistic first Born approximation have been
obtained.

The organization of the paper is as follows. In Sec. II the analytic
expressions for differential cross sections of the SB on an arbitrary
electrostatic potential taking into account the spin interaction as well are
obtained with the help of the dynamic GEA wave function and in the limit of
the Born approximation. In Sec. III we consider the multiphoton cross
sections of SB on screening Coulomb potential.

\section{Multiphoton cross sections of stimulated bremsstrahlung}

The knowledge of the solution of the evolution equation for a Dirac particle
interacting with the electrostatic and electromagnetic fields makes possible
to calculate the scattering amplitude that takes into account the
interaction with the both potential and EM fields simultaneously. When the
wave function describes the particle states only in the region where the
potential energy $U(\overrightarrow{r})$ is not zero then determination of
the scattering amplitude by the asymptote of the wave function \cite{10} is
impossible. Although the scattering amplitude can be defined by the Green
function formalism \cite{11}. As far as the wave function in the GEA
describes the particle states either within the range of the scattering
field or at asymptotic large distances, both approaches with such wave
function are applicable. In this paper we shall consider both approaches
making an emphasis on asymptotic one.

We assume the EM wave to be quasimonochromatic and of an arbitrary
polarization with the vector potential 
\begin{equation}
\label{3abar}\overrightarrow{A}(\varphi )=A_0(\varphi )(\widehat{%
\overrightarrow{e}}_1\cos \varphi +\widehat{\overrightarrow{e}}_2\zeta \sin
\varphi ), 
\end{equation}
where $A_0(\varphi )$ is a slow varying amplitude of the vector-potential of
the plane EM wave $\overrightarrow{A}{\bf (}t,\overrightarrow{r}{\bf )}$
with the phase $\varphi =kx$, $k=(\omega ,\overrightarrow{k})$ is the
four-wave vector of EM field with frequency $\omega $, $\widehat{%
\overrightarrow{e}}_1\cdot \widehat{\overrightarrow{e}}_2=\widehat{%
\overrightarrow{e}}_1\cdot \overrightarrow{k}{\bf =}\widehat{\overrightarrow{%
e}}_2\cdot \overrightarrow{k}=0,$ and $\left| \widehat{\overrightarrow{e}}%
_1\right| =\left| \widehat{\overrightarrow{e}}_2\right| =1$, and $\arctan
\zeta $ is the polarization angle.

The state of the particle in EM wave field is characterized by the average
four- kinetic momentum ( ''quasimomentum'') $\Pi =(\Pi _0,\overrightarrow{%
\Pi })$ defining via free electron four-momentum $p=(\varepsilon _f,%
\overrightarrow{p})$ and relativistic invariant parameter of the wave
intensity $Z$ by the following equation

\begin{equation}
\label{3bar}\Pi =p+kZ(1+\zeta ^2);\qquad Z=\frac{e^2\overline{A}_0^2}{%
4k\cdot p}, 
\end{equation}
($\overline{A}_0$ is the averaged value of the $A_0(\varphi )$) with
corresponding Volkov wave function 
\begin{equation}
\label{4}\Psi _{\Pi ,\mu }^V=\frac 1{\sqrt{2\Pi _0}}f_V(\varphi )\exp \left[
iS_V(x)\right] , 
\end{equation}
where

\begin{equation}
\label{5bar}S_V(x)=-\Pi x+\alpha \left( \frac{\overrightarrow{\Pi }}{k\Pi }%
\right) \sin [\varphi -\theta (\overrightarrow{\Pi })]-\frac{Z(1-\zeta ^2)}2%
\sin 2\varphi 
\end{equation}
is the classical action of charged particle in EM wave field (\ref{3abar})
and 
\begin{equation}
\label{6bar}f_V(\varphi )=u_p^\mu -\frac{eA_0(\varphi )(\gamma k)}{2(k\Pi )}%
\left[ (\overrightarrow{\gamma }\cdot \overrightarrow{e}_1)\cos \varphi
+\zeta (\overrightarrow{\gamma }\cdot \overrightarrow{e}_2)\sin \varphi
)\right] u_p^\mu . 
\end{equation}
is the bispinor amplitude, where $\gamma =(\gamma _0,\overrightarrow{\gamma }%
)$ are the Dirac matrices, $u_p^\mu $ is the bispinor amplitude of a free
particle with polarization $\mu $ and four-momentum $p$ and mass $m$ , $%
\overline{u}_p^\mu u_p^\mu =2m$ , $\overline{u}_p=u_p^{\dagger }\gamma _0$.
Here the quantities $\alpha (\overrightarrow{\rho })$ , $\theta (%
\overrightarrow{\rho })$ as a functions of any vector $\overrightarrow{\rho }
$ are defined by the relations 
\begin{equation}
\label{7bar}\alpha (\overrightarrow{\rho })=e\overline{A}_0\sqrt{\left( 
\overrightarrow{\rho }\cdot \widehat{\overrightarrow{e}}_1\right) ^2+\zeta
^2\left( \overrightarrow{\rho }\cdot \widehat{\overrightarrow{e}}_2\right) ^2%
} 
\end{equation}
and 
\begin{equation}
\label{8bar}\theta (\overrightarrow{\rho })=\arctan \left( \frac{%
\overrightarrow{\rho }\cdot \widehat{\overrightarrow{e}}_2}{\overrightarrow{%
\rho }{\bf \cdot }\widehat{\overrightarrow{e}}_1}\zeta \right) . 
\end{equation}
where $e$ is the particle charge, and the products like $\rho x$, $k\rho $, $%
\gamma k$ are relativistic scalar products: 
$$
\rho x=\rho ^0x^0-\overrightarrow{\rho }{\bf \cdot }\overrightarrow{x}{\bf .}
$$

The wave function of Dirac particle in generalized eikonal approximation
describing induced scattering on an arbitrary electrostatic field in the
presence of strong EM radiation field (\ref{3abar}) has the following form (%
\cite{2})

\begin{equation}
\label{13}\Psi _{\Pi ,\nu }=\frac 1{\sqrt{2\Pi _0}}\left( f_V(\varphi
)+f_1(x)\right) \exp \left[ iS_V(x)+iS_1(x)\right] , 
\end{equation}
where 
$$
S_1(t,\overrightarrow{r})=\frac i{4\pi ^3}\sum\limits_{n=-\infty }^\infty
\int \frac{\widetilde{U}(\overrightarrow{q})\left[ \varepsilon D_n-\omega
\alpha (\frac{\overrightarrow{p}}{kp})D_{1,n}(\theta (\overrightarrow{p}%
))+\omega ZD_{2,n})\right] }{\overrightarrow{q}^2+2\overrightarrow{p}{\bf %
\cdot }\overrightarrow{q}{\bf +}2Z\overrightarrow{k}\cdot \overrightarrow{q}%
-2n(k\cdot p-\overrightarrow{k}{\bf \cdot }\overrightarrow{q})-i0} 
$$
\begin{equation}
\label{14}\times \exp \left( i\left\{ -n\varphi +\overrightarrow{q}\cdot 
\overrightarrow{r}+\alpha _1(\overrightarrow{q})\sin \left[ \varphi -\theta
_1(\overrightarrow{q})\right] -\alpha _2(\overrightarrow{q})\sin 2\varphi
+\theta _1(\overrightarrow{q})n\right\} \right) d\overrightarrow{q}, 
\end{equation}
and 
\begin{equation}
\label{15}f_1(t,\overrightarrow{r})=\frac 1{(2\pi )^3}\sum\limits_{n=-\infty
}^\infty \int \frac{F_n(\varphi ,\overrightarrow{q})d\overrightarrow{q}}{%
\overrightarrow{q}^2+2\overrightarrow{p}{\bf \cdot }\overrightarrow{q}{\bf +}%
2Z\overrightarrow{k}\cdot \overrightarrow{q}-2n(k\cdot p-\overrightarrow{k}%
{\bf \cdot }\overrightarrow{q})-i0} 
\end{equation}
are the action and bispinor amplitude which describe the impact of both the
scattering and EM radiation fields on the particle state simultaneously.
Here 
$$
F_n(\varphi ,\overrightarrow{q})=\widetilde{U}(\overrightarrow{q})\exp
\left( i\left\{ -n\varphi +\overrightarrow{q}{\bf \cdot }\overrightarrow{r}%
+\alpha _1(\overrightarrow{q})\sin \left[ \varphi -\theta _1(\overrightarrow{%
q})\right] -\alpha _2(\overrightarrow{q})\sin 2\varphi +\theta _1(%
\overrightarrow{q})n\right\} \right) 
$$
$$
\times \left\{ D_n(\overrightarrow{\gamma }{\bf \cdot }\overrightarrow{q}%
)\gamma _0+ZD_{2,n}\frac{(\overrightarrow{k}{\bf \cdot }\overrightarrow{q}%
)(\gamma \cdot k)\gamma _0-\omega (\gamma \cdot k)(\overrightarrow{\gamma }%
{\bf \cdot }\overrightarrow{q})}{k\cdot p-\overrightarrow{k}\cdot 
\overrightarrow{q}}+\frac{(\gamma \cdot k)(\overrightarrow{\gamma }{\bf %
\cdot }\overrightarrow{D})(\overrightarrow{\gamma }{\bf \cdot }%
\overrightarrow{q})\gamma _0}{2(k\cdot p-\overrightarrow{k}\cdot 
\overrightarrow{q}{\bf )}}\right. 
$$
$$
+\frac 1{2k\cdot p}\left[ \frac{\omega \left[ \overrightarrow{q}^2+2%
\overrightarrow{p}{\bf \cdot }\overrightarrow{q}{\bf -}2\frac \varepsilon 
\omega \overrightarrow{k}{\bf \cdot }\overrightarrow{q}\right] }{k\cdot p-%
\overrightarrow{k}\cdot \overrightarrow{q}}-(\overrightarrow{\gamma }{\bf %
\cdot }\overrightarrow{q})\gamma _0\right] (\gamma \cdot k)(\overrightarrow{%
\gamma }{\bf \cdot }\overrightarrow{D}) 
$$
$$
+\frac{\omega e\alpha (\frac{\overrightarrow{q}}{kp})(\gamma \cdot k)(%
\overrightarrow{\gamma }{\bf \cdot }\overrightarrow{A}(\varphi ))}{k\cdot p-%
\overrightarrow{k}\cdot \overrightarrow{q}}D_{1,n}(\theta (\overrightarrow{q}%
)) 
$$
\begin{equation}
\label{16}\left. -\frac{e\omega \left[ \overrightarrow{q}^2+2\overrightarrow{%
p}{\bf \cdot }\overrightarrow{q}{\bf -}2\frac \varepsilon \omega 
\overrightarrow{k}{\bf \cdot }\overrightarrow{q}\right] }{2(k\cdot p-%
\overrightarrow{k}\cdot \overrightarrow{q}{\bf )}k\cdot p}(\gamma \cdot k)(%
\overrightarrow{\gamma }{\bf \cdot }\overrightarrow{A}(\varphi ))D_n\right\}
u_p^\nu {\bf ,} 
\end{equation}
where $\widetilde{U}(\overrightarrow{q})=\int U{\bf (}\overrightarrow{\eta }%
)\exp (-i\overrightarrow{q}{\bf \cdot }\overrightarrow{\eta })d%
\overrightarrow{\eta }$ is the Fourier transform of the function $U(%
\overrightarrow{r})$, and $\alpha _1(\overrightarrow{q})$ , $\alpha _2(%
\overrightarrow{q})$ are dynamic parameters of the interaction defining by
expressions 
\begin{equation}
\label{17}\alpha _1(\overrightarrow{q})=\alpha \left( {\bf (}\overrightarrow{%
k}\cdot \overrightarrow{q}{\bf )}\overrightarrow{p}/k\cdot p+\overrightarrow{%
q}\right) ,\ \alpha _2(\overrightarrow{q})=\frac{\overrightarrow{k}\cdot 
\overrightarrow{q}}{2(k\cdot p-\overrightarrow{k}\cdot \overrightarrow{q}%
{\bf )}}Z(1-\zeta ^2), 
\end{equation}
and $\theta _1(\overrightarrow{q})$ is the phase angle 
\begin{equation}
\label{18}\theta _1(\overrightarrow{q})=\theta \left( {\bf (}\overrightarrow{%
k}\cdot \overrightarrow{q}{\bf )}\overrightarrow{p}/k\cdot p+\overrightarrow{%
q}\right) . 
\end{equation}
The functions $D_n$, $D_{1,n}(\theta (\overrightarrow{p}))$, $D_{1,n}(\theta
(\overrightarrow{q}))$ and $D_{2,n}$ are defined by relations 
\begin{equation}
\label{21}D_n=J_n(\alpha _1(\overrightarrow{q}),-\alpha _2(\overrightarrow{q}%
),\theta _1(\overrightarrow{q})), 
\end{equation}
$$
D_{1,n}(\theta (\overrightarrow{p}))=\frac 12\left[ J_{n-1}(\alpha _1(%
\overrightarrow{q}),-\alpha _2(\overrightarrow{q}),\theta _1(\overrightarrow{%
q}))e^{-i\left( \theta _1(\overrightarrow{q})-\theta (\overrightarrow{p}%
)\right) }\right. 
$$
\begin{equation}
\label{22}\left. +J_{n+1}(\alpha _1(\overrightarrow{q}),-\alpha _2(%
\overrightarrow{q}),\theta _1(\overrightarrow{q}))e^{i\left( \theta _1(%
\overrightarrow{q})-\theta (\overrightarrow{p})\right) }\right] , 
\end{equation}
$$
D_{2,n}=(1+\zeta ^2)D_n+\frac{(1-\zeta ^2)}2\left[ J_{n-2}(\alpha _1(%
\overrightarrow{q}),-\alpha _2(\overrightarrow{q}),\theta _1(\overrightarrow{%
q}))e^{-i2\theta _1(\overrightarrow{q})}\right. , 
$$
\begin{equation}
\label{23}\left. +J_{n+2}(\alpha _1(\overrightarrow{q}),-\alpha _2(%
\overrightarrow{q}),\theta _1(\overrightarrow{q}))e^{i2\theta _1(%
\overrightarrow{q})}\right] , 
\end{equation}
$$
\overrightarrow{D}\equiv e\overline{A}_0\left\{ \frac{\widehat{%
\overrightarrow{e}}_1+i\zeta \widehat{\overrightarrow{e}}_2}2J_{n-1}(\alpha
_1(\overrightarrow{q}),-\alpha _2(\overrightarrow{q}),\theta _1(%
\overrightarrow{q}))e^{-i\theta _1(\overrightarrow{q})}\right. 
$$
\begin{equation}
\label{24}+\left. \frac{\widehat{\overrightarrow{e}}_1-i\zeta \widehat{%
\overrightarrow{e}}_2}2J_{n+1}(\alpha _1(\overrightarrow{q}),-\alpha _2(%
\overrightarrow{q}),\theta _1(\overrightarrow{q}))e^{i\theta _1(%
\overrightarrow{q})}\right\} . 
\end{equation}

In the denominator of the integrals in the expression (\ref{14}) and (\ref
{15}) $-i0$ is an imaginary infinitesimal, which shows how the path around
the pole in the integrand should be chosen to obtain a certain asymptotic
behavior of the wave function, i.e. the outgoing spherical wave. Note that
the wave functions is normalized for the one particle in the unit volume.

Let us determine the scattering amplitude by the Green function formalism in
GEA (for the elastic scattering see \cite{1}). For the transition amplitude
in the EM wave field from the state with the ''quasimomentum'' $\Pi $ and
the polarization $\nu $ to the state with the ''quasimomentum'' $\Pi
^{\prime }$ and the polarization $\mu $ we have the expression 
\begin{equation}
\label{24a}T^{\mu \nu }(\Pi \rightarrow \Pi ^{\prime })=\int \overline{\Psi }%
_{\Pi ^{^{\prime }},\mu }^V(x)\gamma _0\Psi _{\Pi ,\nu }(x)U(\overrightarrow{%
r})d^4x 
\end{equation}
where $x$ is the four-radius vector, $\overline{\Psi }=\Psi ^{\dagger
}\gamma _0$, $\Psi ^{\dagger }$ denotes the transposition and complex
conjugation of $\Psi $. According to (\ref{4}) and (\ref{13}) the transition
amplitude (\ref{24a}) can be expressed in the following form

\begin{equation}
\label{24aa}T^{\mu \nu }(\Pi \rightarrow \Pi ^{\prime })=\int e^{i(\Pi
_0^{\prime }-\Pi _0)t}B(t,\overrightarrow{r})dt, 
\end{equation}
where 
\begin{equation}
\label{24b}B(t,\overrightarrow{r})=\int e^{-i(\Pi _0^{\prime }-\Pi _0)t}%
\overline{\Psi }_{\Pi ^{^{\prime }},\mu }^V(x)\gamma _0\Psi _{\Pi ,\nu }(x)U(%
\overrightarrow{r})d\overrightarrow{r} 
\end{equation}
is the periodic function of time. So making a Fourier transformation of the
function $B(t,\overrightarrow{r})$ over $t$ by the relations 
\begin{equation}
\label{24c}B(t,\overrightarrow{r})=\sum_{n=-\infty }^\infty \widetilde{B_n}%
\exp (-int){\bf ,} 
\end{equation}
\begin{equation}
\label{24d}\widetilde{B_n}=\frac \omega {2\pi }\int_{-\pi /\omega }^{\pi
/\omega }B(t,\overrightarrow{r})\exp (int)dt{\bf ,} 
\end{equation}
and carrying out the integration over $t$ in the formula (\ref{24aa}) we
obtain 
\begin{equation}
\label{24e}T^{\mu \nu }(\Pi \rightarrow \Pi ^{\prime })=2\pi \widetilde{B_n}%
\delta \left( \Pi _0^{^{\prime }}-\Pi _0-n\omega \right) . 
\end{equation}
The differential probability of SB process per unit time in the phase space $%
d\overrightarrow{\Pi }^{\prime }/\left( 2\pi \right) ^3$ (space volume $V=1$
in accordance with normalization of electron wave function) is 
\begin{equation}
\label{30}dW_{\Pi \rightarrow \Pi ^{\prime }}=\lim_{t\rightarrow \infty }%
\frac 1t\left| T^{\mu \nu }(\Pi \rightarrow \Pi ^{\prime })\right| ^2\left| 
\overrightarrow{\Pi }^{\prime }\right| \Pi _0^{\prime }d\Pi _0^{\prime }%
\frac{d\Omega }{(2\pi )^3} 
\end{equation}
where $d\Omega $ is the differential solid angle.

Dividing the differential probability of SB process $dW_{\Pi \rightarrow \Pi
^{\prime }}$ by initial flux density $\left| \overrightarrow{\Pi }\right|
/\Pi _0$ and summing over the particle final states and averaging over
initial polarization states, and integrating over $\Pi _0^{\prime }$ we
obtain the differential cross section of SB process for the non-polarized
particles 
\begin{equation}
\label{25}\frac{d\sigma }{d\Omega }=\sum_n\frac{d\sigma ^{(n)}}{d\Omega }, 
\end{equation}
where

\begin{equation}
\label{26}\frac{d\sigma ^{(n)}}{d\Omega }=\left. \frac 1{32\pi ^2}\frac{%
\left| \overrightarrow{\Pi }^{\prime }\right| }{\left| \overrightarrow{\Pi }%
\right| }\sum\limits_{\mu \nu }\left| \widetilde{B_n}\right| ^2\right| _{\Pi
_0^{^{\prime }}=\Pi _0+n\omega } 
\end{equation}
is the partial differential cross section which describes $n$- photon SB
process. Because of very complicated analytical expressions for multiphoton
cross sections of SB in considering approximation (GEA) the ultimate results
require numerical investigations which will be presented elsewhere.

Now let us proceed to the asymptotic approach to construct the multiphoton
cross sections of SB. Note that at asymptotic large distances $r\rightarrow
+\infty $ the GEA wave function coincides with the Born approximation one
when $\left| S_1\left( \overrightarrow{r}{\bf ,}t\right) \right| \ll 1$ (\ref
{15}). As far as we consider an inelastic scattering the wave function of
the particle at large distances $r\rightarrow +\infty $ will be the sum of
spherical convergent waves with the superposition of a plane wave 
\begin{equation}
\label{1bar}\lim_{r\rightarrow +\infty }\Psi (\overrightarrow{r},t)=u_p^\nu
e^{i\overrightarrow{p}{\bf \cdot }\overrightarrow{r}-i\varepsilon
_0t}+\sum_{n=-\infty }^\infty G_n(\widehat{\overrightarrow{r}})\frac{%
e^{i\left| \overrightarrow{\Pi }_n\right| r-i\Pi _nt}}r. 
\end{equation}
where $G_n(\widehat{\overrightarrow{r}})$ is a bispinor depending on $%
\widehat{\overrightarrow{r}}=\overrightarrow{r}/r$. Each term of sum
describes $n$ photon SB process and the partial inelastic scattering
amplitude will be defined as 
\begin{equation}
\label{2bar}f_n^{\mu \nu }(\overrightarrow{\Pi }\rightarrow \overrightarrow{%
\Pi }^{\prime })=\frac 1{2m}\overline{u}_{p^{\prime }}^\mu G_n(\widehat{%
\overrightarrow{r}}), 
\end{equation}
and for the partial differential cross section of SB for the non-polarized
particles we have: 
\begin{equation}
\label{2abar}\frac{d\sigma ^{(n)}}{d\Omega }=\frac 12\frac{\left| 
\overrightarrow{\Pi }^{^{\prime }}\right| }{\left| \overrightarrow{\Pi }%
\right| }\sum\limits_{\mu \nu }\left| f_n^{\mu \nu }(\overrightarrow{\Pi }%
\rightarrow \overrightarrow{\Pi }^{\prime })\right| ^2 
\end{equation}

As the wave function of the particle for SB process in GEA (\ref{15}) in
asymptotic limit of large $r$ has the form 
$$
\lim_{r\rightarrow +\infty }\Psi (\overrightarrow{r},t)=\lim_{r\rightarrow
+\infty }\frac 1{\sqrt{2\Pi _0}}\exp \left( iS_V(x)\right) 
$$
\begin{equation}
\label{4bar}\times \left\{ f_V(\varphi )+\frac{\exp \left[ -i\overrightarrow{%
\Pi }\cdot \overrightarrow{r}\right] }{4\pi r}\sum\limits_{n\ =\ n_0}^\infty
e^{i(\Pi _n\widehat{\overrightarrow{r}}{\bf -}n\overrightarrow{k}{\bf )}%
\cdot \overrightarrow{r}}F_n\left( \varphi ,\overrightarrow{q_n}\right)
\right\} , 
\end{equation}
where$\overrightarrow{q}_n=\Pi _n\widehat{\overrightarrow{r}}-%
\overrightarrow{\Pi }{\bf -}n\overrightarrow{k}$, 
\begin{equation}
\Pi _n=\sqrt{\overrightarrow{\Pi }^2+n\omega \left( 2\Pi _0+n\omega \right) }%
, 
\end{equation}
$F_n\left( \varphi ,\overrightarrow{q_n}\right) $defines by relation (\ref
{16}) at the $r\rightarrow +\infty $ and $\overrightarrow{q}_n=\Pi _n%
\widehat{\overrightarrow{r}}-\overrightarrow{\Pi }{\bf -}n\overrightarrow{k}%
, $ then from the Eqs. (\ref{2bar}), (\ref{4bar}) follows that the bispinor $%
G_n(\widehat{\overrightarrow{r}})$ is the function $F_n(r\rightarrow +\infty
,\overrightarrow{q})$ (the unessential phase corrections are neglected): 
$$
G_n(\widehat{\overrightarrow{r}})=F_n(r\rightarrow +\infty ,\overrightarrow{q%
}_n)=\frac 1{4\pi }\widetilde{U}(\overrightarrow{q}_n) 
$$
$$
\times \left\{ D_n(\overrightarrow{\gamma }{\bf \cdot }\overrightarrow{q}%
_n)\gamma _0+ZD_{2,n}\frac{(\overrightarrow{k}{\bf \cdot }\overrightarrow{q}%
_n)(\gamma k)\gamma _0-\omega (\gamma k)(\overrightarrow{\gamma }{\bf \cdot }%
\overrightarrow{q}_n)}{kp-\overrightarrow{k}\cdot \overrightarrow{q}_n}+%
\frac{(\gamma k)(\overrightarrow{\gamma }{\bf \cdot }\overrightarrow{D})(%
\overrightarrow{\gamma }{\bf \cdot }\overrightarrow{q}_n)\gamma _0}{2(kp-%
\overrightarrow{k}\cdot \overrightarrow{q}_n{\bf )}}\right. 
$$
$$
+\frac 1{2k\cdot p}\left[ \frac{\omega \left[ \overrightarrow{q}_n^2+2%
\overrightarrow{p}{\bf \cdot }\overrightarrow{q}_n{\bf -}2\frac \varepsilon 
\omega \overrightarrow{k}{\bf \cdot }\overrightarrow{q}_n\right] }{kp-%
\overrightarrow{k}\cdot \overrightarrow{q}_n}-(\overrightarrow{\gamma }{\bf %
\cdot }\overrightarrow{q}_n)\gamma _0\right] (\gamma k)(\overrightarrow{%
\gamma }{\bf \cdot }\overrightarrow{D}) 
$$
\begin{equation}
\label{9bar}\left. -2\left[ \varepsilon D_n-\omega \alpha \left( \frac{%
\overrightarrow{p}}{kp}\right) D_{1,n}(\theta (\overrightarrow{p}))+\omega
ZD_{2,n}\right] \right\} u_p^\nu , 
\end{equation}
Here the functions $D_n,D_{1,n},D_{2,n}$ and $\overrightarrow{D}$ are
defined by the expressions (\ref{21})-(\ref{24}), and 
\begin{equation}
\label{10bar}\alpha _1(\overrightarrow{q}_n)=\alpha _1\left( \frac{%
\overrightarrow{p}^{\prime }}{kp^{\prime }}-\frac{\overrightarrow{p}}{kp}%
\right) ,\ \alpha _2(\overrightarrow{q}_n)=\frac{Z^{\prime }-Z}2(1-\zeta
^2), 
\end{equation}
\begin{equation}
\label{11bar}\theta _1(\overrightarrow{q}_n)=\theta \left( \frac{%
\overrightarrow{p}^{\prime }}{kp^{\prime }}-\frac{\overrightarrow{p}}{kp}%
\right) =\theta \left( \frac{\overrightarrow{\Pi }^{\prime }}{k\Pi ^{\prime }%
}-\frac{\overrightarrow{\Pi }}{k\Pi }\right) . 
\end{equation}

In addition, taking into account that $\overline{u}_{p^{\prime
}}(p_0^{\prime }\gamma _0-\overrightarrow{\gamma }\overrightarrow{p}^{\prime
}-m)=0$ and $(p_0\gamma _0-\overrightarrow{\gamma }\overrightarrow{p}%
-m)u_p=0 $ and known relations between the $\gamma $-matrix elements 
$$
\gamma ^\mu \gamma ^\nu -\gamma ^\nu \gamma ^\mu =2\delta ^{\mu \nu } 
$$
we reduce the transition amplitude to the following form 
\begin{equation}
\label{18bar}f_n^{\mu \nu }(\overrightarrow{\Pi }\rightarrow \overrightarrow{%
\Pi }^{\prime })=-\frac 1{4\pi }\overline{u}_{p^{\prime }}^\mu Au_p^\nu 
\widetilde{U}(\overrightarrow{q}_n), 
\end{equation}
where 
\begin{equation}
\label{20bar}A=\widehat{E}+\widehat{k}^{\prime }\widehat{D}, 
\end{equation}
$$
\widehat{E}=\gamma _0D_n+\frac{\omega Z}{kp^{\prime }}(\gamma k)D_{2,n}\text{
,} 
$$
\begin{equation}
\label{21bar}~k^{\prime }=\left( \frac k{kp}+\frac{\widetilde{k}}{kp^{\prime
}}\right) /2,~\widetilde{k}=\left( k_0,-\overrightarrow{k}\right) . 
\end{equation}

Then introducing (\ref{18bar}) into the (\ref{2abar}) the scattering cross
section will be write in the form

\begin{equation}
\label{19bar}\frac{d\sigma ^{(n)}}{d\Omega }=\frac{\left| \overrightarrow{%
\Pi }^{^{\prime }}\right| \left| \widetilde{U}(\overrightarrow{q}_n)\right|
^2}{(4\pi )^2\left| \overrightarrow{\Pi }\right| }2Sp\left\{ \rho ^{\prime
}A\rho \overline{A}\right\} , 
\end{equation}
where $~$%
$$
\rho =\frac{\widehat{p}+m}2,~\rho ^{\prime }=\frac{\widehat{p}^{\prime }+m}2 
$$
are the initial and final density matrixes.

Taking into account that $kp=k\Pi ,$ $kp^{\prime }=k\Pi ^{\prime }$ and
using the properties of defined functions $D_n,D_{1,n},D_{2,n}$ and $%
\overrightarrow{D}$ (which follow directly from Eqs. (\ref{21})-(\ref{24})
and relation (\ref{A8})) we obtain the following expression for the partial
differential cross sections of SB process 
$$
\frac{d\sigma ^{(n)}}{d\Omega }=\frac{\left| \overrightarrow{\Pi }^{^{\prime
}}\right| }{(4\pi )^2\left| \overrightarrow{\Pi }\right| }\left| \widetilde{U%
}(\overrightarrow{q_n})\right| ^2\left\{ 4\left| \varepsilon D_n+\omega
ZD_{2,n}-\omega \alpha \left( \frac{\overrightarrow{\Pi }}{k\Pi }\right)
D_{1,n}(\theta (\overrightarrow{\Pi }))\right| ^2\right. 
$$
\begin{equation}
\label{32bar}\left. -\overrightarrow{q}_n^2\left| D_n\right| ^2+\frac 1{%
(k\Pi )(k\Pi ^{\prime })}[\omega ^2\overrightarrow{q}_n^2-(\overrightarrow{k}%
\cdot \overrightarrow{q}_n)^2]\left( \left| \overrightarrow{D}\right| ^2-%
\frac{e^2\overline{A}_0^2}2ReD_nD_{2,n}^{\dagger }\right) \right\} , 
\end{equation}
where 
$$
\left| \overrightarrow{D}\right| ^2=\frac{e^2\overline{A}_0^2}4\left[
(1+\zeta ^2)\left( \left| J_{n-1}\right| ^2+\left| J_{n+1}\right| ^2\right)
+2(1-\zeta ^2)\left[ \cos 2\theta _1(\overrightarrow{q}_n)ReJ_{n-1}J_{n+1}^{%
\dagger }\right. \right. . 
$$
\begin{equation}
\label{33bar}\left. +2i\sin 2\left( \theta _1(\overrightarrow{q}_n)-\theta (%
\overrightarrow{p})\right) Im\left( J_{n-1}J_{n+1}^{\dagger }\right) \right]
. 
\end{equation}
Comparing the cross sections of SB process for spinor and scalar particles
we can conclude that the spin interaction is described by the terms in order
of square of the quantum recoil \symbol{126}$\overrightarrow{q}_n^2$ and
gives a considerable contribution in the SB cross sections only for the
large-angle scattering (which is known also for the elastic scattering from
the formula of Mott ) and for the relativistic intensities of EM wave at $%
K=eA/m\geq 1$.

\section{Differential cross sections of SB on the screening Coulomb
potential for the circular and linear polarizations of EM wave}

For concreteness we utilize the Eq. \ref{32bar} to obtain the differential
cross section of SB on a screening Coulomb potential for which the Fourier
transform is 
\begin{equation}
\label{33abar}\widetilde{U}(\overrightarrow{q}_n{\bf )=}\frac{4\pi Z_ae^2}{%
\overrightarrow{q}_n^2+\chi ^2}{\bf ,} 
\end{equation}
where $1/\chi $ is the radius of screening, $Z_a$ is the charge number of
the nucleus.

For circular polarized EM wave the quantities in (\ref{32bar}) are given by
relations (\ref{8bar})-(\ref{11bar}), (\ref{21})-(\ref{24}) at $\zeta =1$.
Then taking into account that at $\alpha _2(\overrightarrow{q})=0$ the
functions $D_n$, $\,D_{1,n}(\theta (\overrightarrow{\Pi }))$ and $D_{2,n}$
are expressed by ordinary Bessel functions, for the partial differential
cross section of SB we have 
$$
\frac{d\sigma ^{(n)}}{d\Omega }=\frac{(Z_ae^2)^2\left| \overrightarrow{\Pi }%
^{^{\prime }}\right| }{\left| \overrightarrow{\Pi }\right| \left( 
\overrightarrow{q}_n^2+\chi ^2\right) ^2}\left\{ J_n^2(\alpha _1(%
\overrightarrow{q}_n))\left[ 4\left( \Pi _0-\frac{n\omega \alpha (\frac{%
\overrightarrow{\Pi }}{k\Pi })}{\alpha _1(\overrightarrow{q}_n)}\cos \left[
\theta _1(\overrightarrow{q}_n)-\theta (\overrightarrow{p})\right] \right)
^2\right. \right. 
$$
\begin{equation}
\label{35bar}\left. -\overrightarrow{q}_n^2+\left. \beta ^2\left( \frac{n^2}{%
\alpha _1^2(\overrightarrow{q}_n)}-1\right) \right] +J_n^{^{\prime
}2}(\alpha _1(\overrightarrow{q}_n))\left( 4\omega ^2\alpha ^2(\frac{%
\overrightarrow{p}}{kp})\sin ^2\left[ \theta _1(\overrightarrow{q}_n)-\theta
(\overrightarrow{p})\right] +\beta ^2\right) \right\} , 
\end{equation}
where 
\begin{equation}
\alpha _1(\overrightarrow{q}_n)=\alpha _1\left( \frac{\overrightarrow{\Pi }%
^{\prime }}{k\Pi ^{\prime }}-\frac{\overrightarrow{\Pi }}{k\Pi }\right) , 
\end{equation}
\begin{equation}
\theta _1(\overrightarrow{q}_n)=\theta _1\left( \frac{\overrightarrow{\Pi }%
^{\prime }}{k\Pi ^{\prime }}-\frac{\overrightarrow{\Pi }}{k\Pi }\right) 
\end{equation}
\begin{equation}
\beta ^2=\frac{e^2\overline{A}_0^2}{(kp)(kp^{\prime })}[\omega ^2%
\overrightarrow{q}_n^2-(\overrightarrow{k}{\bf \cdot }\overrightarrow{q}%
_n)^2]. 
\end{equation}
and $J_n^{^{\prime }}(\vartheta )$ denotes the first derivative of ordinary
Bessel function with respect to $\vartheta .$

In the case of linearly polarized EM wave all quantities are defined for $%
\zeta =0$. As far as $\theta _1(\overrightarrow{q}_n)=\theta (%
\overrightarrow{p})=0$ , the functions $D_n$, $\,D_{1,n}(\theta (%
\overrightarrow{\Pi }))$ and $D_{2,n}$ are defined by the real function $%
J_n(u,v)$ \cite{8}, further called the generalized Bessel function \cite{9}.
Making the simple transformations and using the expression \ref{A8} 
$$
2nJ_n(u,v)=u\left[ J_{n-1}(u,v)+J_{n+1}(u,v)\right] +2v\left[
J_{n-2}(u,v)+J_{n+2}(u,v)\right] 
$$
we obtain the partial differential cross section of SB process $d\sigma
^{(n)}/d\Omega $ for the linear polarization of the wave 
$$
\frac{d\sigma ^{(n)}}{d\Omega }=\frac{(Z_ae^2m)^2\left| \overrightarrow{\Pi }%
^{^{\prime }}\right| }{\left| \overrightarrow{\Pi }\right| \left| 
\overrightarrow{q}_n^2+\chi ^2\right| ^2}\left\{ J_n^2(u,v)\left[ \epsilon
^2-\overrightarrow{q}_n^2-\beta ^2\left( \frac 12+\frac n{4v}\right) \right]
\right. 
$$
$$
+I_n^2(u,v)\left[ 4\omega ^2\alpha ^{\prime 2}+\beta ^2\right] 
$$
\begin{equation}
\label{36bar}\left. +J_n(u,v)I_n(u,v)\left[ \frac{u\beta ^2}{4v}-4\omega
\epsilon \alpha ^{\prime }\right] \right\} , 
\end{equation}
where 
$$
I_n(u,v)=\frac 12\left( J_{n-1}(u,v)+J_{n+1}(u,v)\right) , 
$$
and 
$$
\epsilon =2\Pi _0+\frac{n\omega Z}v 
$$
$$
\alpha ^{\prime }=\alpha (\frac{\overrightarrow{\Pi }}{k\cdot \Pi })+\frac{uZ%
}{2v}. 
$$
The arguments $u$, $v$ are determined by the relations 
$$
u=e\overline{\overrightarrow{A}}_0\cdot \left( \frac{\overrightarrow{\Pi }%
^{\prime }}{k\Pi ^{\prime }}-\frac{\overrightarrow{\Pi }}{k\Pi }\right) , 
$$
$$
v=-\alpha _2(\overrightarrow{q}_n)=\frac{Z-Z^{\prime }}2. 
$$

Comparing the non relativistic cross section \cite{3} with relativistic one
it is easy to see that besides the additional terms, which come from
spin-orbital and spin-laser interaction ($\sim \overrightarrow{q}_n^2$) as
well as effect of intensity ($\sim K^2$), the relativistic contribution is
conditioned by arguments of the Bessel functions. Because of sensitivity of
the Bessel function to relationship of it's argument and index the most
probable number of emitted or absorbed photons will be defined by the
condition $\left| n\right| \sim \left| \alpha _1(\overrightarrow{q}%
_n)\right| $. By this reason the contribution of relativistic effects on the
scattering process, as have been shown in Ref. (\cite{7}), becomes essential
already for $K\sim 0.1,$ consequently, the dipole approximation is violated
for nonrelativistic parameters of interaction..

In Fig.1 the numerical calculations are presented for the same parameters as
were in Ref. (\cite{7}) as for the circular as well as for linear
polarization of EM wave. In Fig.1a. the envelopes of partial differential
cross sections as a function of the number of emitted or absorbed photons
are shown for the deflection angle $\angle \overrightarrow{\Pi }%
\overrightarrow{\Pi }^{\prime }=0.6$ $mrad$, for an intensity of Neodymium
laser $3.5\times 10^{16}W/cm^2$ ( which corresponds to relativistic
parameter of intensity $K\simeq 0.17$) and a moderate initial electron
kinetic energy $\varepsilon _k=2.7$ keV. The blue and green curves
correspond to initial electron momentum parallel and antiparallel to the
laser propagation direction $\overrightarrow{k}$ respectively and the red
curve gives the nonrelativistic result. Note, that in case of colinear $%
\overrightarrow{k}$ and $\overrightarrow{p}$ there is an azimutal symmetry
with respect to propagation direction.

In Fig.1b. the envelopes of partial differential cross sections for linear
polarization of EM wave are shown. To emphasize the differences the
relativistic parameter of intensity and initial electron kinetic energy are
taken the same. In this case there is no azimutal symmetry and we have taken
the final electron momentum in the same plane with $\overrightarrow{A}$ and $%
\overrightarrow{k}$. The deflection angle is again $\angle \overrightarrow{%
\Pi }\overrightarrow{\Pi ^{\prime }}=0.6mrad$.

The energy exchange increases for large deflection angles and for the high
intensities of EM wave. In Fig.2 the envelopes of partial differential cross
sections as a function of the number of emitted or absorbed photons for
circular polarization of EM wave are shown for the deflection angle $\angle 
\overrightarrow{\Pi }\overrightarrow{\Pi ^{\prime }}=6mrad.$ The laser
parameters and initial electron kinetic energy are taken the same. As is
seen from Fig.2, the differences between the cases of initial electron
momentum parallel or antiparallel to the laser propagation direction $%
\overrightarrow{k}$on the one hand and between nonrelativistic result on the
other hand are notable.

To show the dependence of SB process upon laser intensity in the Fig.3 the
total differential cross sections are plotted as a function of relativistic
parameter of intensity $K$. Fig.3a. and Fig.3b. correspond to initial
electron momentum parallel and antiparallel to the laser propagation
direction $\overrightarrow{k},$ respectively, and the Fig.3 gives the
nonrelativistic result.

\begin{center}
{\bf Acknowledgments}
\end{center}

We would like to thank Prof. H.K. Avetissian for valuable discussions during
the working under the present paper, scientific seminars of Plasma Physics
Laboratory of Yerevan State University leading by him are gratefully
acknowledged.

\appendix

\section{Definition of the Function $J_n(u,v,\triangle )$}

A function $J_n(u,v,\triangle )$ may be defined by the expression 
\begin{equation}
\label{A1}J_n(u,v,\triangle )=(2\pi )^{-1}\int_{-\pi }^\pi d\theta \exp
\left[ i\left( u\sin (\theta +\triangle )+v\sin 2\theta -n(\theta +\triangle
)\right) \right] 
\end{equation}
or by an infinite series representation 
\begin{equation}
\label{A2}J_n(u,v,\triangle )=\sum_{k=-\infty }^\infty e^{-i2k\triangle
}J_{n-2k}(u)J_k(v). 
\end{equation}
Both defining relations are equivalent. From either Eq. (\ref{A1}) or (\ref
{A2}) it follows that 
\begin{equation}
\label{A3}J_n(u,0,\triangle )=J_n(u), 
\end{equation}
and 
\begin{equation}
\label{A4}J_n(0,v,\triangle )=\left\{ 
\begin{array}{c}
e^{-i\triangle n}J_{
\frac n2}(v),\ n\text{ even} \\ 0,\ n\text{ odd} 
\end{array}
\right. \ . 
\end{equation}
Then we have directly relative formulas 
$$
J_n(-u,v,\triangle )=(-1)^nJ_n(u,v,\triangle ), 
$$
$$
J_n(u,-v,\triangle )=(-1)^nJ_{-n}(u,v,-\triangle ), 
$$
\begin{equation}
\label{A5}J_n(u,-v,-\triangle )=(-1)^nJ_{-n}(u,-v,\triangle ). 
\end{equation}
From the well known recurrence relations for the Bessel functions we have 
\begin{equation}
\label{A6}J_{n-1}(u,v,\triangle )-J_{n+1}(u,v,\triangle )=2\partial
_uJ_n(u,v,\triangle ), 
\end{equation}
and 
\begin{equation}
\label{A7}e^{-i2\triangle }J_{n-2}(u,v,\triangle )-e^{i2\triangle
}J_{n+2}(u,v,\triangle )=2\partial _vJ_n(u,v,\triangle ), 
\end{equation}
that follows directly from Eq. (\ref{A1}) or (\ref{A2}).

The integration by parts in Eq. (\ref{A1}) yields to the following relation 
$$
2nJ_n(u,v,\triangle )=u\left[ J_{n-1}(u,v,\triangle )+J_{n+1}(u,v,\triangle
)\right] 
$$
\begin{equation}
\label{A8}+2v\left[ e^{-i2\triangle }J_{n-2}(u,v,\triangle )+e^{i2\triangle
}J_{n+2}(u,v,\triangle )\right] . 
\end{equation}
Other results can be obtained by combination of Eqs. (\ref{A2})-(\ref{A8}).
We perform two important formulas, which can be proved from Eq. (\ref{A1}).
The first is 
\begin{equation}
\label{A9}\sum_{n=-\infty }^\infty e^{in(\varphi +\triangle
)}J_n(u,v,\triangle )=\exp \left\{ i\left[ u\sin (\varphi +\triangle )+v\sin
2\varphi \right] \right\} , 
\end{equation}
and the second is 
\begin{equation}
\label{A10}\sum_{k=-\infty }^\infty J_{n\mp k}(u,v,\triangle )J_k(u^{\prime
},v^{\prime },\pm \triangle )=J_n(u\pm u^{\prime },v\pm v^{\prime
},\triangle ). 
\end{equation}
Then the function $J_n(u,v,\triangle )$ at $\triangle =0$ turns to the
generalized Bessel function $J_n(u,v),$ that was induced by Reiss in Ref.
[12].

{\bf Figure captions}

Figure 1. The envelopes of partial differential cross sections $d\sigma
^{(n)}/d\Omega $ in atomic units as a function of the number of emitted or
absorbed photons, for an intensity of Neodymium laser $3.5\times
10^{16}W/cm^2$ , $\omega =1.17eV$. The radius of screening is$1/\chi =4$
a.u.,$Z_a=1$, the initial electron kinetic energy $\varepsilon _k=2.7$ keV:
a) and b) correspond to circular and linear polarization of EM wave
respectively. The deflection angle equals $\angle \overrightarrow{\Pi }%
\overrightarrow{\Pi ^{\prime }}=0.6$ mrad. The blue and green curves
correspond to initial electron momentum parallel and antiparallel to the
laser propagation direction $\overrightarrow{k}$ respectively and the red
curve gives the nonrelativistic result.

Figure 2. The same as Fig.1, for the deflection angle $\angle 
\overrightarrow{\Pi }\overrightarrow{\Pi ^{\prime }}=6$ mrad.

Figure 3. The total differential cross sections $d\sigma /d\Omega $ are
plotted as a function of relativistic parameter of intensity $K$ in the
range $0<K<1$. Fig.3a. and Fig.3b. correspond to initial electron momentum
parallel and antiparallel to the laser propagation direction $%
\overrightarrow{k},$ respectively, and the Fig.3c. gives the nonrelativistic
result.

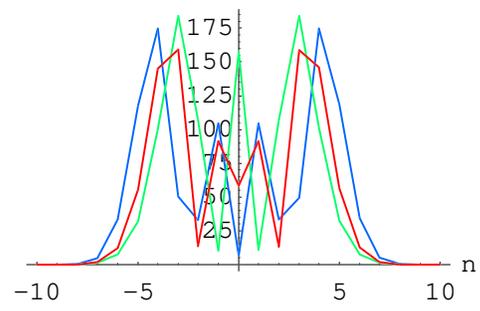
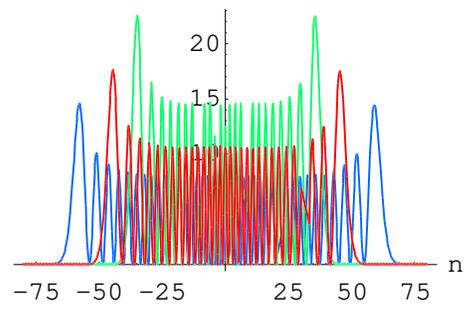
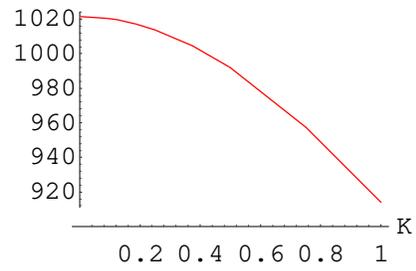
\end{document}